\documentclass[letterpaper,twoside,12pt]{article}
\usepackage{amssymb}
\usepackage{amsmath}
\usepackage{latexsym}
\usepackage{verbatim}
\usepackage{graphicx}
\usepackage{epsfig}
\usepackage{pstricks}
\usepackage{stmaryrd}
\usepackage{rotating,graphicx}
\usepackage[english]{babel}
\usepackage{setspace}
\usepackage[english]{babel}
\usepackage[utf8]{inputenc}
\usepackage{natbib}
\usepackage[T1]{fontenc}
\usepackage{lmodern}
\usepackage{enumitem}
\usepackage{natbib}

\begin{document}

\newcommand{\bra}[1]{\langle #1|}
\newcommand{\ket}[1]{|#1\rangle}
\newcommand{\braket}[2]{\langle #1|#2\rangle}

\begin{Large}
\begin{center}
\textbf{Black Holes, Information Loss and the Measurement Problem}\\
\end{center}
\end{Large}

\begin{center}
\begin{large}
Elias Okon\\
\end{large}
\textit{Instituto de Investigaciones Filos\'oficas, Universidad Nacional Aut\'onoma de M\'exico, Mexico City, Mexico.}\\[.5cm]
\begin{large}
Daniel Sudarsky\\
\end{large}
\textit{Instituto de Ciencias Nucleares, Universidad Nacional Aut\'onoma de M\'exico, Mexico City, Mexico.} \\
\end{center}

\noindent
The \emph{information loss paradox} is often presented as an unavoidable consequence of well-established physics. However, in order for a genuine paradox to ensue, not-trivial assumptions about, e.g., quantum effects on spacetime, are necessary. In this work we will be explicit about these additional, speculative assumptions required. We will also sketch a map of the available routes to tackle the issue, highlighting the, often overlooked, commitments demanded of each alternative. In particular, we will display the strong link between black holes, the issue of information loss and the measurement problem.

\onehalfspacing

\section{Introduction}
The so-called \emph{information loss paradox} is usually introduced as an unavoidable consequence of standard, well-established physics. The paradox is supposed to arise from a glaring conflict between Hawking's black hole radiation and the fact that time evolution in quantum mechanics preserves information. However, the truth is that, in order for a genuine paradox to appear, a sizable number of additional, non-standard assumptions is required. As we will see, these extra assumptions involve thesis regarding the fundamental nature of Hawking's radiation, guesses regarding quantum aspects of gravity and even considerations in the foundations of quantum theory.

In this work, we will be explicit about the additional assumptions required for a genuine conflict to arise and delineate the available options in order to tackle the issue. In particular, we will stress the connection between information loss and the measurement problem, and display the often non-trivial commitments that each of the available alternatives to solve the information loss issue demands.
\section{The classical setting: black holes hide information}
\label{C}
We start by reviewing some properties of classical black holes. Gravity, being always attractive, tends to draw matter together to form clusters. In fact, if the mass of a cluster is big enough, nothing will be able to stop the contraction until, eventually, a black hole will form. That is, the gravitational field at the surface of the body will be so strong that not even light will be able to escape and a region of spacetime from which nothing is able to emerge will form. The boundary of such a region is called the event horizon and, according to general relativity, its area never decreases. 

In general, the collapse dynamics that leads to the formation of a black hole can, of course, be very complicated. However, it can be shown that all such systems eventually settle down into one of the few stationary black hole solutions, which are completely characterized by the mass, charge and angular momentum of the the Kerr-Newman spacetimes. In fact, the so-called black hole uniqueness theorems guarantee that, as long as one only considers gravitational and electromagnetic fields, then these solutions represent the complete class of stationary black holes. Moreover, the so-called no-hair theorems ensure that the set of stationary solutions does not grow, even if one considers other hypothetical fields.

The above mentioned results seem to suggest that when a cluster collapses to form a black hole, a large amount of information is lost. That is, details such as the multipole moments of the initial mass distribution, or the type of matter involved, seem to be altogether lost when the black hole settles. Note however that such apparent loss of information corresponds only to that available to observers outside of the black hole. While at early times there are Cauchy hypersurfaces\footnote{A Cauchy hypersurface is a subset of spacetime which is intersected exactly once by every inextensible, non-spacelike curve.} completely contained outside of the black hole, at later times all Cauchy hypersurfaces have parts both inside and outside it (see Figure 1). Therefore, using data located both outside and inside of the black hole, the \emph{whole} spacetime can always be recovered. We conclude that, in the classical setting, information is not really lost. All that happens is that, when a black hole forms, a new region of no escape emerges and some of the information from the outside of the black hole moves into such new region. One could still argue that, since there are points inside of the horizon which are not in the past of future null infinity,\footnote{Future null infinity is the set of points which are approached asymptotically by null rays which can escape to infinity.}  then it is impossible to reconstruct the whole spacetime by evolving backwards the data on it. However, future null infinity is not a Cauchy hypersurface so one should not expect to reconstruct the whole spacetime from such data.

\begin{figure}[h]
\centering
 \begin{pspicture}(4.5,5)
 \psline(1,0)(1,4)(2.25,4)(3.5,2.5)(1,0)
\pscustom[linewidth=.5pt,fillstyle=solid,fillcolor=gray]{
\pscurve(1,0)(1.4,2)(1.3,4)
\psline[liftpen=1](1,4)(1,0)}
\psline[linewidth=2pt](1,4)(2.25,4)
\psline[linewidth=.5pt](1,2.75)(2.25,4)
\rput(2.6,1){$\mathcal{I}^-$}
\rput(3.25,3.5){$\mathcal{I}^+$}
\rput(0,3.5){\small Horizon}
\psline{->}(.7,3.47)(1.65,3.47)
\rput(1.6,4.5){\small Singularity}
\psline{->}(1.6,4.35)(1.6,4.05)
\rput(-1,2){\small Collapsing body}
\psline{->}(.42,2)(1.2,2)
 \end{pspicture}
\caption{Penrose diagram for a collapsing spherical body. $\mathcal{I}^+$ and $\mathcal{I}^-$ denote past and future null infinity.}
\end{figure}
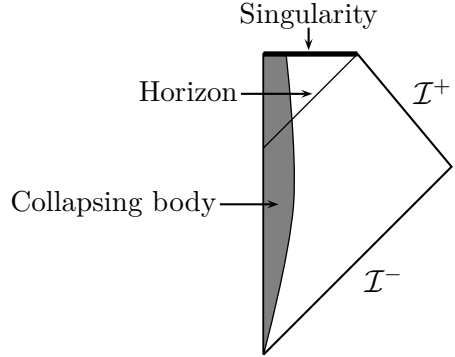

\section{QFT on a fixed curved background: black holes radiate}
\label{H}
The most dramatic change in our understanding of black hole physics came as a result of Hawking's famous analysis. 
What this analysis showed was that the formation of a black hole would modify the state of any quantum field in such a way that, at late times, there would be an outgoing flux of particles carrying energy towards infinity. Moreover, Hawking showed that the flux was characterized by the surface gravity $\kappa$ of the resulting asymptotic stationary state of the black hole. This discovery transformed our perception of the formal analogy, originally pointed out in \cite{Bekenstein}, between the laws of black hole dynamics, and the standard laws of thermodynamics (see \cite{Wald} for a discussion). In particular, it led to the view that the surface gravity is in fact a measure of the black hole's temperature $T=\frac{\kappa}{2\pi}$, and that the event horizon's area $A$ is a measure of the black hole's entropy $S=A/4$. 
 
Hawking's result is probably the most famous of the effects that arise from the natural extension of special relativistic quantum field theory to the realm of curved spacetimes. It imposes a dramatic modification on the classical view of black holes as absolutely black and eternal regions of spacetime. It is important to stress, though, that Hawking's calculation, being a result pertaining to quantum field theory on a \emph{fixed} spacetime, does not encompass back-reaction effects. These are in fact notoriously difficult to deal with and a general framework for doing so is lacking. At any rate, some straightforward physical considerations, which have rather dramatic consequences, are often brought to bear in this context. 
 
\section{Back-reaction and first quantum gravity input: black holes evaporate}
\label{E}
As can be expected, Hawking's result also suggests a dramatic modification in our expectation for the ultimate fate of a black hole. That is, while before Hawking's discovery, one would have expected that, once formed, a black hole would be eternal, the fact that the radiation is caring energy away, assuming overall energy conservation, leads one to expect that the mass of the black hole will start diminishing. The context in which this problem is standardly set is that of asymptotically flat spacetimes, for which we have a well defined notion of overall energy content given by the ADM mass\footnote {The ADM mass is a quantity associated with the asymptotic behavior of the induced spatial metric of a Cauchy hypersurface. In asymptotically flat spacetimes, it is known to be independent of the hypersurface on which it is evaluated (see \cite{ADM}).} of the spacetime, a quantity which is known to be conserved.
 
As we noted, Hawking's calculation cannot deal with back-reaction. However, our confidence on energy conservation in the appropriate situations is so robust that it is difficult not to conclude that, as the radiation carries away energy, the black hole mass will have to diminishing. If this takes place, the surface gravity of the black hole---which is no longer really stationary, but can be expected to deviate from stationarity only to a very small degree---would change as well. As it turns out, the surface gravity is inversely proportional to the black hole's mass, so the black hole temperature can be expected to increase, leading to a ever more rapid rate of energy loss and a correspondingly faster decrease in mass. 

The run away picture for the evaporation process suggests a complete disappearance of the black hole in a finite amount of time. Of course, we cannot really be sure about this picture because, in order to perform a solid analysis, we would need to deploy a, currently lacking, trustworthy theoretical formalism adept to the challenge. 
The problem is that, by the removal of energy from the black hole, one can expect to eventually reach a regime where quantum aspects of gravitation become essential to the description of the process. 
At such point, one might contemplate the possibility that, as a result of purely quantum gravitational aspects, the Hawking evaporation of the black hole will stop, leaving a small stable remnant. This, in turn, might open certain possibilities regarding the information issue. For the time being, though, we will ignore such an option. 

Then, in order to simplify the discussion at this point, we will ignore the possibility of remnants and assume that there is nothing to stop the Hawking radiation. Then, if the black hole's mass decreases in accordance with energy conservation, one expects that the black hole to simply disappear and the spacetime region where it was located to turn flat (see Figure 2).

\begin{figure}[h]
\centering
 \begin{pspicture}(4.5,5)
 \psline(1,0)(1,3)(2,3)(2,4)(3.5,2.5)(1,0)
\pscustom[linewidth=.5pt,fillstyle=solid,fillcolor=gray]{
\pscurve(1,0)(1.4,2)(1.35,3)
\psline[liftpen=1](1,3)(1,0)}
\psline[linewidth=2pt](1,3)(2,3)
\psline[linewidth=.5pt](1,2)(2,3)
\rput(2.6,1){$\mathcal{I}^-$}
\rput(3.25,3.5){$\mathcal{I}^+$}
\rput(0,2.66){\small Horizon}
\psline{->}(.7,2.62)(1.57,2.62)
\rput(.85,4.5){\small Singularity}
\psline{->}(.85,4.35)(1.5,3.1)
\rput(-1,1.5){\small Collapsing body}
\psline{->}(.42,1.5)(1.21,1.5)
 \end{pspicture}
\caption{Penrose diagram for a collapsing spherical body, taking into account Hawking's radiation.}
\end{figure}
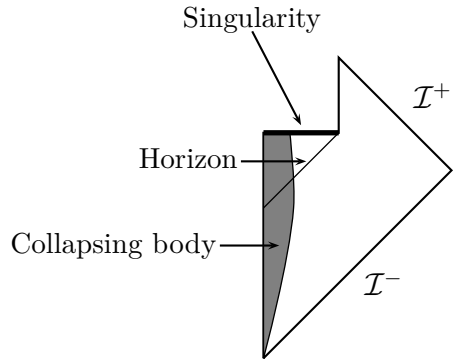
 
At this point, we seem to come face to face with an information loss problem: the original massive object that collapses, leading to the formation of a black hole, might have required an incredibly large amount of detail for its description. However, the final state that results from the evaporation is simply described in terms of the thermal Hawking flux, followed by an empty region of spacetime. 
More to the point, even if the initial matter that collapses to form a black hole was initially in a pure quantum state, after the complete evaporation of the black hole there would be a mixed one, corresponding to the thermal Hawking flux. These considerations seem to indicate that, even at the fundamental level, we have a fundamental loss of information. The final state, even if described in full detail, does not encode the information required to retrodict the details of the initial one. At the level of quantum theory, we would be facing a non-unitary (and non-deterministic) relation between the initial and final states of the system, a situation that seems at odds with the unitary evolution provided by the Schrödinger equation.

There are, however, various caveats to the above conclusion. The first one is opened up by the possibility of the evaporation eventually stopping, leading to a stable remnant. The mass of said remnant can be estimated by considering the natural scales at which the effects of quantum gravity are expected to become important. This leads to an estimate of the order of Plank's mass ($ \approx 10^{-5}$ gr). Then, if one wants the remnant to encode all the information present in the initial state, one is led to the conclusion that such a small object would have a number of possible internal states as large as that of the original matter that collapsed to form the black hole, which can, of course, have had a mass as large as one can imagine. 
It is hard, then, to envisage what kind of object, with such rather unusual thermodynamical behavior, would this remnant have to be. For this reason, this possibility is usually not considered viable (although we acknowledge that these considerations might be overturned; for a discussion of these issues see \cite{Remnants}). At any rate, we will not consider this possibility any further. 

We should also mention another proposal which uses the idea that, while curing singularities, quantum gravity might open paths to other universes, which could be home to the missing information. Such information would be encoded either in a new universe or in correlations between it and ours. Besides the dramatic ontological burden, such proposal leaves open the possibility of these alternative universes emerging even in ordinary processes (which could, e.g., involve virtual black holes), leading to information loss in such standard scenarios. Alternatively, the information could be preserved, but impossible to retrieve in principle. We will also not consider this possibility any further. 

A much more important caveat is the following: we have very solid results indicating that, associated with the formation of a black hole, there is always a singularity of spacetime appearing withing it. The strongest results in this regard are a series of theorems proved by Hawking (see \cite{Hawking-Thms}) showing that, under quite general conditions, and assuming reasonable properties for the energy and momentum of the collapsing matter, the formation of singularities is an inevitable result of Einstein's equations. The issue is that, at the classical level, these singularities represent a breakdown of the theory and, in fact, a failure of the spacetime description. The singularities are, therefore, to be thought of as representing boundaries of spacetime, rather that points within it. Once a spacetime has additional boundaries, it is clear that the issue of information has to be confronted on a different light. Of course, if one considers the description of the system at an initial Cauchy hypersurface and wants a final hypersurface to encode the same information, one has to make sure that the final one is also Cauchy.

The formation of singularities then implies that, if we want to have spacetime regions where the system's state could be thought of as encoding all the information, then we must surround the singularities by suitable boundaries. In other words, if the singularities force us to include further boundaries of spacetime, then the comparison of initial and final information has to be done between the initial Cauchy hypersurface and the late-time \emph{collection} of surfaces that, together, act as a Cauchy hypersurface. That collection could naturally include asymptotically null future, but also the hypersurfaces surrounding the singularities. The same kind of calculation as the one done by Hawking would then show that all the information present on the initial hypersurface would also be encoded in the state associated with this late-time Cauchy hypersurface. That is, if we include the boundary of spacetime that arises in association with the singularity, then there is no issue regarding the fate of information. 
We conclude that, under these circumstances, still there is no information loss.

\section{Second quantum gravity input: black holes do not involve singularities}
\label{S}
As we noted above, singularities represent a breakdown of the spacetime description as provided by general relativity and thus indicate the need to go beyond such theory. The expectation among theorists is that quantum gravity is going to be the theory that cures these failures of classical general relativity, replacing the singularities by a description in the language appropriate to quantum gravity. This is, in fact, what occurs with various other theories that are known to be just effective descriptions of a physical system's behavior in a limited context, but that have to be replaced with a more fundamental description once the system leaves that regime. Think for instance of the description of a fluid by, say, the Navier-Stokes equations. We know that this description works very well in a large variety of circumstances, but that a breakdown of such description occurs, for instance, when there are shock waves or when other types of singularities are formed. However, under such circumstances, the underlying kinetic theory, including the complex inter-molecular forces, is expected to remain valid. 
The point is that, just as in those cases, one expects the emergence of singularities in general relativity to indicate the end of the regime where the classical description of spacetime is valid and, therefore, where a quantum gravity description would have to take over (see Figure 3 and \cite{Ash} for details). 
 
\begin{figure}[h]
\centering
 \begin{pspicture}(4.5,5)
 \psline(1,0)(1,5)(3.5,2.5)(1,0)
\pscustom[linewidth=.5pt,fillstyle=solid,fillcolor=gray]{
\pscurve(1,2.3)(1.75,2.3)(2.5,2.5)(1.75,2.7)(1,2.7)}
\psline[linewidth=.5pt,linestyle=dashed](1,1)(2.5,2.5)
\psline[linewidth=.5pt](2.5,2.5)(3,3)
\psline[linewidth=.5pt,linestyle=dashed](1,4)(3,2)
\rput(2.6,1){$\mathcal{I}^-$}
\rput(2.6,4){$\mathcal{I}^+$}
\pscurve[linewidth=.5pt](1,2.5)(2,2)(2.5,2.5)
 \end{pspicture}
\caption{``Quantum spacetime diagram'' for a black hole.}
\end{figure}
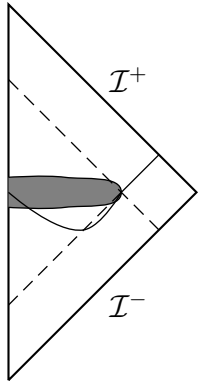 
 
Of course, if quantum gravity does in fact cure the singularities, and removes the need to consider, in association with the corresponding regions, a boundary of spacetime, the issue of the fate of information in the Hawking evaporation of black holes resurfaces with dramatic force. So, do we finally have a genuine paradox in our hands. Not quite yet; a few elements are still missing. In order for a paradox to arise, we need to couple a genuine loss of information with a fundamental theory which does not allow for information to be lost.

\section{A paradox?}
\label{P}
When is it, then, that the Hawking radiation by a black hole leads to an actual paradox? We are finally in a position to enumerate the various assumptions required in order to construct a genuine conflict:
\begin{enumerate}
\item As a result of Hawking's radiation carrying energy away from the black hole, the mass of the black hole decreases and it either evaporates completely or leaves a small remnant.
\item In the case where the black hole leaves a small remnant, the number of its internal degrees of freedom is bounded by its mass in such a way that these cannot possibly encode the information contained in an arbitrarily massive initial state.
\item Information is not transfered to a parallel universe.
\item As a result of quantum gravity effects, the internal singularities within black holes are cured and replaced by something that eliminates the need to consider internal boundaries of spacetime.
\item The outgoing radiation does not encode the initial information.
\item Quantum evolution is always unitary.
\end{enumerate}
 
We have already discussed the arguments in support of assumptions 1, 2, 3 and 4 and saw that, although by no means conclusive, they are reasonable. But what about 5 and 6? Well, in order to avoid a paradox, and assuming the first four assumptions to be true, at least one of them has to be negated. In order to explore the motivations and consequences of doing so, we must think clearly about how to interpret Hawking's calculation in a context in which 1, 2, 3 and 4 are the case.

As we remarked above, Hawking's calculation is performed in the setting of a quantum field theory over a fixed curved background. What one finds there is that an initial pure state of the field evolves into a final one which, when tracing over the inside region, reduces to a mixed thermal state. The key question at this point, then, is how to interpret such a final mixed state in a setting in which i) the black hole is no longer there, so there is no interior region to trace over, and ii) in which there is no singularity (or parallel universe) for the information to ``escape into.'' As far as we can see, there are two alternatives: either one assumes that the mixed state arises only as a result of tracing over the interior region and maintains that the outgoing radiation somehow encodes the initial information---which amounts to negating 5; or one takes Hawking's result seriously and maintains that, even in this scenario, information is lost---which amounts to negating 6. Below we explore each option in detail.

\subsection{The outgoing radiation encodes information}
In the last couple of decades, the community's position on the information loss subject has been strongly influenced by developments in String theory. Such framework has permitted exploration of questions, regarding black holes, using settings where event horizons and singularities play no relevant roles. This is possible due to the AdS/CFT correspondence (see e.g., \cite{AdS/CFT}), which allows the mapping of complicated spacetime geometries in the ``bulk'' of asymptotically Anti-de Sitter spacetimes, including ones involving black holes, onto corresponding states of an ordinary quantum field theory living on the Anti-de Sitter boundary (which is, in fact, a flat spacetime). These considerations have led people to conclude that, as a breakdown of unitarity is not expected to take place in the context of a quantum field theory in flat spacetimes, there should be no room for a breakdown of unitarity in the corresponding situation involving black holes either.
\footnote{Note however that the argument can be easily reversed to show exactly the opposite. Since Hawking's result shows that unitarity breaks when black holes are present, one must conclude that quantum evolution \emph{cannot} be unitary even in a quantum field theory on flat spacetimes.} 

The proposal, then, is that unitarity is never broken and that information is never lost. As a result, Hawking's calculation has to be somehow attuned to assure consistency. In particular, the proposal is that the outgoing radiation must encoded all of the initial information. There is, however, a high price to pay in order to achieve this. As has been shown in \cite{FireWalls}, in order for the outgoing radiation to encode the necessary information, each emitted particle must get entangled with all the radiation emitted before it. However, due to the so-called, ``monogamy of entanglement,'' doing so entails the release of an enormous amount of energy, turning the event horizon into a \emph{firewall} that burns anything falling through it. The upshot then, is a divergence of the energy-momentum tensor of the field over the event horizon and a radical breakdown of the equivalence principle over such a region.

\subsection{Unitarity is broken}
The discovery of the Hawking radiation was initially taken as a clear indicative of information loss at the fundamental level. In fact, \cite{HawkingOp} even introduced a notation for this general type of evolution which was supposed to account for the transformation from (possibly pure) initial states $\rho_i $ into final mixed ones $\rho_f $. Hawking denoted the general linear, non-unitary, operator characterizing such transformation by the sign $ \$$, i.e., $ \rho_f = \$ \rho_i $. Likewise, Penrose pointed out that, in order to have a consistent picture of phase space for situations involving black holes in thermal equilibrium with an environment, one has to assume that ordinary quantum systems undergo something akin to a self-measurement, by which he meant quantum state reduction that was not the result of measurement by external observers or measuring devices (see \cite{Penrose-BHequilibrium}). \cite{Penrose-Emperor's} further argued that quantum state reduction is probably linked to aspects of quantum gravity. 
 
The early assessments of these ideas in \cite{Preskil} indicated that they where likely to lead to a very serious conflict with energy and momentum conservation or to generate unacceptable non-local features in ordinary physical situations. However, further analysis in \cite{Wald-Unruh} showed that these assessments where not that solid and that there where various possibilities to evade the apparently damning conclusions.
 
In (omitted references) 
we have explored the viability of breaking unitarity both qualitatively and quantitatively. In particular, we have successfully adapted objective collapse models, developed in connection with foundational issues within quantum theory, in order to explicitly describe the transition from the initial pure state into a mixed one. Our view on the subject is based on the conviction that, contrary to the prevailing opinion in the community working on the gravity/quantum interface, there are good reasons to think that quantum theory requires modifications to deal with its basic conceptual difficulties. Below we discuss these issues and explore their consequences for the information loss paradox.

\section{Information loss and the measurement problem}
\label{MP}
Most discussions of black holes and information loss do not implicate foundational issues of quantum theory. Of course, ignoring such issues, particularly with pragmatic interests in mind, is often acceptable. However, when deep conceptual questions are involved, such as in the present case, the pragmatic attitude might not be the right way to go. 

The standard interpretation of quantum mechanics involves a profoundly \emph{instrumentalist} character, with notions such as \emph{observer} or \emph{measurement} playing a crucial role. Such an instrumentalist trait becomes a problem as soon as one intends to regard the theory as a fundamental one, useful not only to make predictions in suitable experimental settings, but also to be applied to the measurement apparatuses, to the observers involved, or to non-standard contexts such as black holes or the universe as a whole. The resulting problem, often referred to as the \emph{measurement problem}, has been discussed at length in numerous places and many different concrete formulations of it have been given. A particularly useful way to state it, given in \citet{Maudlin}, is as a list of three statements that cannot be all true at the same time:
\begin{description}[font=\normalfont,labelindent=.5cm]
\item[A.] The physical description given by the quantum state is complete.
\item[B.] Quantum evolution is always unitary.
\item[C.] Measurements always yield definite results.
\end{description}

Maudlin's formulation of the measurement problem is noteworthy because of its generality and its preciseness. Moreover, it is extremely useful in order to motivate and classify strategies to solve the problem. For example, by negating A, one arrives at so-called hidden variable theories, such as Bohmian mechanics; by removing B, one gets so-called objective collapse theories, such as GRW; and by discarding C, Everettian interpretations emerge. Of these three options, the last one is, by far, the most contentious. Among its most urgent matters, we can mention the problem of the preferred basis, the one of making sense of probabilities in the theory and the general and basic issue of establishing a clear and precise link between the abstract mathematical objects of the theory and concrete empirical predictions. Of course, brave attempts to deal with these and other issues within Everettian frameworks abound. However, be believe that, at least for the time being, they are far from being successful.

Returning to the measurement problem and its relation to the information loss issue, we note that assumptions 6 and B are in fact identical. Therefore, the strategy one decides to adopt in order to avoid complications regarding the information loss issue (e.g., negating 5 or 6 above) has implications with respect to what one must say regarding the measurement problem (e.g., negating A, B or C). In particular, if regarding the information loss, one decides to maintain the validity of 6 (and thus to hold that the outgoing radiation encodes all of the initial information), then one necessarily has to either negate A or C (i.e., either to entertain a hidden variables theory or an Everettian scenario). In other words, insisting on a purely unitary evolution, not only demands a violation of the equivalence principle and a divergence of the energy-momentum tensor, but also a commitment either with many worlds or with an acknowledgment that standard quantum mechanics is incomplete. On the other hand, if regarding the information loss problem, one decides to abandon unitarity, the same move automatically not only avoids a breakdown of the equivalence principle, but also guarantees success with respect to the measurement problem. The upper hand of the second option seems evident to us.

\section{Conclusions}
\label{Con}
Since the publication of Hawking's analysis, more than forty years ago, the issue of black hole information loss has been a central topic in theoretical physics. The AdS/CFT correspondence, proposed almost twenty years latter, came to further propel an already notorious debate. Yet, even after all these years, the discussion is often engulfed by confusion and misunderstanding among participants. The objective of this work is to develop a clear analysis of some of the key conceptual issues involved. Our hope is that, by doing so, significant progress on this important topic could soon be achieved.

We have presented the basic theoretical setting of the black hole information issue, paying special attention to elements, arising from not yet well-established physics, that presently have to be regarded merely as reasonable assumptions. Moreover, we have argued that the information loss issue is closely related to the measurement problem, and claimed that it is precisely within the context of certain proposals put forward to deal with the latter that the former finds one of its most conservative resolutions.

\bibliographystyle{apalike}
\bibliography{PSA16.bib}

\begin{thebibliography}{}

\bibitem[Almheiri et~al., 2013]{FireWalls}
Almheiri, A., Marolf, D., Polchinski, J., and Sully, J. (2013).
\newblock Black holes: complementarity or firewalls?
\newblock {\em JHEP}, 62.

\bibitem[Arnowitt et~al., 1962]{ADM}
Arnowitt, R., Deser, S., and Misner, C. (1962).
\newblock The dynamics of general relativity.
\newblock In Witten, L., editor, {\em Gravitation: an introduction to current
  research}. Wiley.

\bibitem[Ashtekar and Bojowald, 2005]{Ash}
Ashtekar, A. and Bojowald, M. (2005).
\newblock Black hole evaporation: a paradigm.
\newblock {\em Class. Quant. Grav.}, 22(3349).

\bibitem[Banks, 1994]{Remnants}
Banks, T. (1994).
\newblock Lectures on black hole information loss.
\newblock {\em Nucl. Phys. Proc.}, 41.

\bibitem[Banks et~al., 1984]{Preskil}
Banks, T., Susskind, L., and Preskin, M.~E. (1984).
\newblock Difficulties for the evolution of pure states unto mixed states.
\newblock {\em Nucl. Phys. B}, 224(125).

\bibitem[Bekenstein, 1972]{Bekenstein}
Bekenstein, J.~D. (1972).
\newblock Black holes and the second law.
\newblock {\em Lett. Nuovo Cim.}, 4(737).

\bibitem[Hawking, 1976]{HawkingOp}
Hawking, S.~W. (1976).
\newblock Breakdown of predictability in gravitational collapse.
\newblock {\em Phys. Rev. D}, 14(2460).

\bibitem[Hawking and Ellis, 1973]{Hawking-Thms}
Hawking, S.~W. and Ellis, G. F.~R. (1973).
\newblock {\em The large scale structure of spacetime}.
\newblock Cambridge University Press.

\bibitem[Maudlin, 1995]{Maudlin}
Maudlin, T. (1995).
\newblock Three measurement problems.
\newblock {\em Topoi}, 14.

\bibitem[Penrose, 1981]{Penrose-BHequilibrium}
Penrose, R. (1981).
\newblock Time asymmetry and quantum gravity.
\newblock In Isham, C.~J., Penrose, R., and Sciama, D.~W., editors, {\em
  Quantum Gravity II}. Clarendon Press.

\bibitem[Penrose, 1999]{Penrose-Emperor's}
Penrose, R. (1999).
\newblock {\em The Emperor's New Mind: Concerning Computers, Minds, and the
  Laws of Physics}.
\newblock Oxford University Press.

\bibitem[Strominger, 2001]{AdS/CFT}
Strominger, A. (2001).
\newblock The {AdS/CFT} correspondence.
\newblock {\em JHEP}, 0110(034).

\bibitem[Unruh and Wald, 1995]{Wald-Unruh}
Unruh, W.~G. and Wald, R.~M. (1995).
\newblock On evolution laws taking pure states to mixed states in quantum field
  theory.
\newblock {\em Phys. Rev. D}, 52:2176--2182.

\bibitem[Wald, 1994]{Wald}
Wald, R.~M. (1994).
\newblock {\em Quantum Field Theory in Curved Spacetime and Black Hole
  Thermodynamics}.
\newblock University of Chicago Press.

\end{thebibliography}
\end{document}